\documentclass[12pt]{article}
\usepackage{fullpage}

\def\be{\begin{equation}}
\def\ee{\end{equation}}
\def\bea{\begin{eqnarray}}
\def\eea{\end{eqnarray}}

\begin{document}
\begin{flushright}
TIFR/TH/05-39 \\
27.9.2005 
\end{flushright}
\bigskip

\begin{center}
{\LARGE\bf Einstein and the Quantum} \\[1cm]
{\Large Virendra Singh} \\
Tata Institute of Fundamental Research \\
Homi Bhabha Road, Mumbai 400 005, India \\[2cm]
\underbar{\bf Abstract}
\end{center}

We review here the main contributions of Einstein to the quantum
theory.  To put them in perspective we first give an account of
Physics as it was before him.  It is followed by a brief account of
the problem of black body radiation which provided the context for
Planck to introduce the idea of quantum.  Einstein's revolutionary
paper of 1905 on light-quantum hypothesis is then described as well as
an application of this idea to the photoelectric effect.  We next take
up a discussion of Einstein's other contributions to old quantum
theory.  These include (i) his theory of specific heat of solids,
which was the first application of quantum theory to matter, (ii) his
discovery of wave-particle duality for light and (iii) Einstein's A
and B coefficients relating to the probabilities of emission and
absorption of light by atomic systems and his discovery of radiation
stimulated emission of light which provides the basis for laser
action.  We then describe Einstein's contribution to quantum
statistics viz Bose-Einstein Statistics and his prediction of
Bose-Einstein condensation of a boson gas.  Einstein played a
pivotal role in the discovery of Quantum mechanics and this is briefly
mentioned.  After 1925 Einstein's contributed mainly to the
foundations of Quantum Mechanics.  We choose to discuss here (i) his
Ensemble (or Statistical) Interpretation of Quantum Mechanics and (ii)
the discovery of Einstein-Podolsky-Rosen (EPR) correlations and the
EPR theorem on the conflict between Einstein-Locality and the
completeness of the formalism of Quantum Mechanics.  We end with some
comments on later developments.

\newpage

\noindent {\Large\bf 1. Physics before Einstein.}
\bigskip

Albert Einstein (1879-1955) is one of the two founders of quantum
theory along with Max Planck.  Planck introduced the `quantum' of
energy in his investigations of black body radiation in 1900.  He was
followed by the young Einstein who proposed the `light quantum
hypothesis' in 1905.  
Albert Einstein sent his revolutionary ``light quantum'' paper for
publication on 17 March 1905 to Annalen der Physik.  He was twenty six
years of age and it was his first paper on quantum theory.  He had
published five papers earlier during 1901-1904 in the same journal.
Those dealt with capillarity and statistical mechanics.  The major
frontier areas of research in physics then were thermodynamics and
electrodynamics.  The main conceptions about the physical universe
prevalent in physics of that time were as follows: 
\bigskip

\noindent {\large 1.1. \underbar{`Newton's mechanical conception'}}
\medskip

The earliest of these was that of a ``mechanical universe'' given by
Isaac Newton in his magnum opus ``Principia'' in 1687.  The physical
universe in it was regarded as composed of discrete point-particle 
endowed with masses.  They moved with time along well defined
trajectories, in the fixed arena of a three dimensional Euclidean space, 
under the influence of mutual forces.  The trajectories could be
deterministically calculated by using Newton's three laws of motion
provided one knew the forces involved and also the initial position
and velocities of all the particles.  The forces involved were of the
``action at a distance'' type.  Newton also discovered the universal
attractive force of gravitation which acts between any two mass points
and falls of as the square of the interparticle distance.  Astronomy
was thereby brought into the fold of physics unlike the case in
Aristotlean physics of ancients.

It was known that there exists other forces such as magnetic forces,
electric forces, chemical affinity etc.  It was part of Post-Newtonian
program of research to determine their laws.  The force law between
two ``magnetic poles'' was determined by John Mitchell in 1750, while
that between two electric charges was conjectured theoretically by
Joseph Priestley, the discoverer of Oxygen, in 1769 and experimentally
verified in the unpublished work of Henry Cavendish done in 1771.  It
was however published first, based on his own work, by Charles Coulomb
in 1785 and is now known as Coulomb's law.  Alessandro Volta used
electric currents, produced by his Voltaic pile, to dissociate a
number of substances e.g. water into Hydrogen and Oxygen.  After this
work it was a clear possibility that the forces responsible for
chemical binding may be reducible to electrical forces.  Matter could
consist entirely of electrically charged mass points.
\bigskip

\noindent {\large 1.2. \underbar{Light as waves}}
\medskip

Newton was also inclined to view light also to be discrete stream of
particles, `light-corpuscles'.  Christian Huygens communicated his
researches on light to members of French Academy in 1678, and
published in 1690 as `Trait\'e de la Lumi\'ere', wherein he advanced
the notion that light is a wave phenomena.  The wave theory of light
got strong boost from the discoveries of interference of light in 1801
by Thomas Young, and by the studies of Augustin Fresnel on diffraction
of light begining in 1815.  As a result the wave theory of light was
firmly established.  It was inconcieable, in those days, to have a
wave motion without a medium for it to propagate, so a ``luminiferous
aether'' was postulated for its' propagation.
\bigskip

\noindent {\large 1.3. \underbar{Energetics program}}
\medskip

We just saw that light had proved refractory to being accomodated within
Newton's mechanical conception of the universe.  In thermodynamics, it
was easy to see that the first law of thermodynamics, which refers to
the law of energy conservation, could be easily interpreted within
Newtonian framework.  However it did not look possible to interpret
the second law of thermodynamics, dealing with increasing entropy,
within it.  Ludwig Boltzmann's H-theorem was an attempt towards this
goal during 1842-1877 using his kinetic theory of gases.  This attempt
attracted strong criticism from Ernst Zermelo and others.  Georg Helm
and Ludwig Ostwald, supported by Ernst Mach, therefore denied the
reality of atoms and suggested that energy is the most fundamental
concept and the whole program of physics should be reduced to a
``generalised thermodynamics''.  This program, ``Energetics'', was
subscribed to by a small but strongly vocal and influencial
minority.  In fact Einstein's work on Brownian motion in 1905 played a
crucial role in it's fall.
\bigskip

\noindent {\large 1.4. \underbar{Electromagnetic conception of the
universe}} 
\medskip

Michael Faraday introduced the concept of continuous fields, like
electric and magnetic fields, defined over the whole space-time,  in
contrast to discrete particles.  He did this in order to have a deeper
understanding of his law of electromagnetic induction in eighteen
thirtees.  These fields are produced by electric charges, and electric
currents produced by these charges in motion.  They then interact with
other electric charges elsewhere.  There is no ``action at a
distance'' but every interaction is a local interaction.  Faraday
quoted the old saying ``matter can not act where it is not'' in a
letter to Richard Taylor in 1844.  Faraday also thought the
gravitational force, which appears to act at a distance between two
masses, could also be understood as a local interaction by the
introduction of a gravitational field.

Clerk Maxwell's equations for electric and magnetic fields, given in
1864, unified these two disparate entities into a coherent single
entity ``electromagnetic field''.  Maxwell, synthesized the earlier
known discoveries of Coulomb's law, Gauss', laws of magnetic
induction, Oersted's work on production of magnetic fields by electric
current, and Faraday's laws of electromagnetic induction into one set
of equations using the field concept.  He also appended a new element,
now called ``Maxwell's displacement current'', to this synthesis.

A brilliant windfall from the Maxwell's equations was the prediction
of the existence of transverse electromagnetic waves with a constant
velocity (now denoted by the latter $c$).  The velocity $c$ agreed
with the known velocity of light.  It was therefore natural for
Maxwell to propose ``electromagnetic wave theory'' of light.  The
subject of optics thus became a branch of electromagnetic theory.  The
luminiferrous aether was identified as the aether for electromagnetic
fields as well.

The tantalising possibility, the electromagnetic conception of the
universe, arose now.  Could it be that even point charged particles
can be viewed as arising from the aether?  The mass of an electron
could be entirely due to it's electromagnetic energy.  If so, the
``electromagnetic aether'' would be the sole ontological entity in
terms of which one would be able to understand the whole nature.
\bigskip

\noindent {\large 1.5. \underbar{Two clouds on the Horizon}}
\medskip

In a lecture delivered in April 1900 before the Royal Institution,
Lord Kelvin talked about two ``Nineteenth Century Clouds Over the
Dynamical Theory of Heat and Light''.  It was such a rare case of
penetrating insight into the nature of physics that one is left
admiring it even now.  It is the resolution of these two ``clouds'' that
gave rise to the two revolutions in twentieth century physics.  One of
these clouds referred to the continued unsuccessful attempts to detect
the motion of the earth through aether and it's resolution was
achieved by Einstein's special theory of Relativity (1905).  We shall
not be dealing with this any further here.  The other cloud referred
to the failure of the equipartition theorem in classical statistical
mechanics.  It resolution required the second revolution, associated
with the quantum.
\bigskip\bigskip

\noindent {\Large\bf 2. The Problem of Blackbody Radiation: From \\
Kirchhoff to Planck}
\medskip

Max Planck, in 1900, was first to introduce the quantum ideas in
physics and he did this in the context of blackbody radiation.  We now
discuss the early history of this problem for providing the setting of
his work.
\medskip

\noindent {\large 2.1. \underbar{Kirchhoff:}}
\medskip

All heated bodies emit and absorb radiation energy.  The emissivity
$e(\lambda,T)$ of a body, for the radiation with wave length
$\lambda$, depends on the nature of body and it's temperature $T$.  It
is the same for it's absorptivity $a(\lambda,T)$.  Using consideration
of thermodynamics equilibrium, it was shown by Gustav Kichhoff of
Berlin, in 1859, that the ratio of emissivity $e(\lambda,T)$ to it's
absorptivity $a(\lambda,T)$ is independent of the nature of the
heated body i.e.
\[
e(\lambda,T) = E(\lambda,T) a(\lambda,T)
\]
where $E(\lambda,T)$ is a universal function of only the wave length
$\lambda$ of the radiation and it's temperature $T$.

If we define, following Kirchhoff, a perfect blackbody as one whose
absorptivity is equal to unity, i.e. perfect absorption, then the
universal function $E(\lambda,T)$ can be identified with the emissivity
of a perfect blackbody.  He also showed that the radiation inside a
heated cavity which is opaque and maintained at temperature $T$,
behaves like blackbody radiation.  One can therefore experimently
study the blackbody radiation by using the radiation issuing out a
cavity through a small hole.
\bigskip

\noindent {\large 2.2. \underbar{Boltzmann:}}
\medskip

Ludwig Boltzmann, in 1884, using Maxwell's electromagnetic theory
showed that 
\[
E(\lambda,T) = (c/8\pi) \rho(\nu,T),
\]
where $\rho (\nu,T)$ is the energy density of radiation at frequency
$\nu$ and temperature $T$.  ($c$ = velocity of light in vacuum, $\nu$
= frequency of the radiation = $c/\lambda$).  He further showed using
thermodynamics consideration, together with Maxwell's relation $P =
{1\over3} u$ between pressure $P$ and energy density $u$ of the
radiation, that the total radiant energy per unit volution is
proportional to $T^4$ i.e.
\[
\int^\infty_0 d\nu \rho(\nu,T) = \sigma T^4
\]
where $\sigma$ is called Stefan-Boltzmann Constant.  Josef 
Stefan had conjectured the truth of this law on the basis of his
experimental work in 1879 for all heated bodies, though it is strictly
true only for a blackbody.
\bigskip

\noindent {\large 2.3. \underbar{Wien:}}
\medskip

Further progress was made by Wilham Wien in 1894, when he studied the
thermodynamics of extremely slow, i.e. adiabatic, contraction of the
cavity on the blackbody radiation contained in it.  From these he
concluded that 
\[
\rho(\nu,T) = \nu^3 f(\nu/T).
\]
This is known as `Wien's displacement law'.  We have thus reduced the
problem of determining $\rho(\nu,T)$, a function of two variables $\nu$
and $T$, to that of determining a function $f(\nu/T)$ of a single
variable $(\nu/T)$.  This is as far as one can go on the basis of
purely thermodynamic considerations.

To give a representation of the experimental data Wien also proposed a
form for this function
\[
\rho(\nu,T) = a \nu^3 e^{-b\nu/T},
\]
which we shall refer to as Wien's radiation law.  In this $a$ and $b$
are numerical coefficients to be fixed from the data.
\bigskip

\noindent {\large 2.4. \underbar{Rayleigh-Jeans:}}
\medskip

In June 1900, Lord Rayleigh decided to apply equipartion theorem of
Maxwell-Boltzmann to the problem of radiation and derived
\[
\rho(\nu,T) = c_1 \nu^2 T.
\]
He did not calculate at that time the numerical coefficient $c_1$, which
he did in May 1905.  He however, made a mistake of a factor of 8 which
was corrected by James Jeans in June 1905.  With the numerical factor
included we have 
\[
\rho(\nu,T) = {8\pi\nu^2 \over c^3} \cdot kT
\]
which is known as Rayleigh-Jeans' radiation law.  Here $k$ is the
Boltzmann constant.  Rayleigh felt that this is a limiting form of
$\rho(\nu,T)$ for $\nu/T \rightarrow 0$.  Note that if this law was
correct for all $\nu$, then it would lead to ultraviolet catastrophe.
The total energy would be infinite.

\newpage

\noindent {\large 2.5. \underbar{Planck:}}
\medskip

Max Planck succeeded to the chair of Kirchhoff at Berlin in 1889.  He
was naturally drawn to the problem of determining the universal
function $\rho(\nu,T)$ introduced by his predecessor.  As he said
``The so called normal energy distribution represents something
absolute, and since the search for absolutes has always appeared to me
to be the highest form of research, I applied myself vigorously to
it's solution''.  He argued that since the universal $\rho(\nu,T)$
does not depend on the nature of the material of walls, it's
determination would be facilitated if one assumes a simple model for
it.  He proposed to regard the wall to be made of Hertzian
oscillators, each one capable of emitting or absorbing radiation of
only a single frequency $\nu$.  He then showed, using electromagnetic
theory i.e.
\[
\rho(\nu,T) = {8\pi\nu^2 \over c^3} {\bar E} (\nu,T)
\]
where $\bar E(\nu,T)$ is the average energy of the Hertzian oscillator
of frequency $\nu$ at temperature $T$.  He had this result on May 18,
1899. 

Earlier experimental work by Friedrich Paschen on blackbody radiation
had shown that Wien's radiation law fitted the data well as it was
known in 1897 for $\lambda = 1 - 8\mu$ and $T = 400 - 1600 \
{^{^\circ}K}$.  Later work by Otto Lummer and Ernst Pringhsheim, in
the region $\lambda = 12 - 18 \mu$ and $T = 300 - 1650 \
{^{^\circ}K}$, had however revealed the deviations from Wien's
radiation law in February 1900.  On Oct 19, 1900 Kurlbaum announced the
measurements done with Rubens for even higher wavelength region,
$\lambda = 30 - 60 \mu$ and $T = 200 - 1500 \ {^{^\circ}K}$. Planck
then gave his radiation law as a discussion remark to this
announcement.  In modern notation, (first done in 1906), it reads as
\[
\rho(\nu,T) = {8\pi\nu^2 \over c^3} \cdot {h\nu \over e^{h\nu/kT} - 1}
\]
where $h$ is now known as Planck's constant.  This suggested radiation
law fitted the data perfectly.  Note also that it reduces to (i)
Rayleigh-Jean's law for $\nu/T \rightarrow 0$ and (ii) has the same
form as Wien's radiation law for $\nu/T \rightarrow \infty$ and (iii)
provides the `correct' interpolation formula between the two regions.
At this stage it was a purely empirical formula without any derivation.
He then got busy looking for one.

Planck, when he began his research career was inclined to
``energetics'' school and believed in the deterministic significance, 
unlike what was advocated by Boltzmann who took the probabilistic
view, of entropy.  In Boltzmann's view the entropy $S$ of a configuration
was related to it's thermodynamic probability $W$ i.e.
\[
S = k \ln W
\]
Planck, as an ``act of desperation'', was forced to use Boltzmann's
view to derive his formula.  In order to calculate thermodynamic
probability for a configuration of $N$ oscillators, with total energy
$U_N = NU$ and entropy $S_N = N S$, he assumed that $U_N$ is made up
of finite energy elements $\epsilon$ i.e. $U_N = P\epsilon$, and
worked out the total number of possible ways $W_N$ of distributing $P$
energy elements $\epsilon$ among $N$ oscillators.  He obtained
\[
W_N = {(N + P - 1)! \over P! (N - 1)!}.
\]
The thermodynamic probability $W$ was taken proportional to $W_N$.
This leads to
\[
S = {S_N \over N} = k\left[\left(1 + {U \over \epsilon}\right)
\ln\left(1 + {U \over \epsilon}\right) - {U \over \epsilon} \ln {U
\over \epsilon}\right].
\]
On using ${\partial S \over \partial U} = {1\over T}$, we obtain
\[
\bar E(\nu,T) = {\epsilon \over e^{\epsilon/kT} - 1},
\]
which on using Wien's displacement law, leads to (in modern notation)
\[
\epsilon = h \nu.
\]
Planck presented this derivation of his radiation law on 14 December
1900 to German Physical Society and this can be taken as the birth
date of quantum theory.  The really new element was his assumption
that the Hertzian oscillators with frequency $\nu$ can emit or absorb
radiation in the units of $\epsilon = h\nu$.  Planck however did not
realise the revolutionary nature of his procedure.  As he said, ``this
was purely a formal assumption and I really did not give it much
thought except that, no matter what the cost, I must bring about a
positive result''.
\bigskip

\noindent {\Large\bf 3. Einstein's Light Quantum paper}
\medskip

\noindent {\large 3.1. \underbar{Light quantum hypothesis:}}
\medskip

Albert Einstein was the first person to have a clear realisation that
Planck's introduction of energy quanta was a revolutionary step and
thus one which would have larger significance for physics than just
for the problem of blackbody radiation.  In 1905, Einstein's annus 
mirabilis, he published his light quantum paper.

Einstein starts in this paper by first noting that the unambiguous
prediction of electrodynamics and equipartition theorem for the
material oscillators is that given by the radiation law, now called
``Rayleigh-Jeans law''.  He is in fact the first person to derive this
law from classical physics correctly as his work was done before Jeans
obtained the proper numerical constant in it.  As such Abram Pais,
even feels that it would be more proper to call it
Rayleigh-Einstein-Jean's law.  Since this radiation law does not agree
with experiments, and theoretically suffers from ``ultraviolet
catastrophe'' (i.e. infinite total energy), it leads to a clear
failure of classical physics.  Something in classical physics has to
yield.

In his search for the cause of failure, Einstein is motivated by his
dissatisfaction with asymmetrical treatment of matter and radiation in
classical physics.  As we saw earlier matter is discrete and
particulate while the radiation is continuous and wave-field like in
classical physics.  He wondered whether the failure of the classical
radiation theory was in not treating radiation also as discrete and
particulate.  He thus proposes his hypothesis of ``light quantum''. Of 
course he is well aware of the enormous success which wave theory of
light had in dealing with the phenomenon of interference, diffraction,
etc. of light.  About this aspect he comments

``The wave theory, operating with continuous spatial functions, has
proved to be correct in representing purely optical phenomena and will
probably not be replaced by any other theory.  One must,  however,
keep in mind that the optical observations are concerned with temporal
mean values and not with instantaneous values, and it is possible, in
spite of the complete experimental verification of the theory of
reflection, refraction, diffraction, dispersion and so on that the
theory of light which operates with continuous spatial functions may
lead to contradictions with observations if we apply it to the
phenomenon of generation and transformation of light''.

Einstein then proceeds to show that an analysis of ``experimental''
Wien's radiation law, valid in ``nonclassical'' regime of large
$\nu/T$, gives an indication of the particle nature.  For this purpose
he does an elaborate calculation of the probability $p$ that the
monochromatic radiation of frequency $\nu$, occupying a volume $V_0$,
could all be found later in a volume $V$.  He finds this, on using
Wien's radiation law, to be given by
\[
p = \left(V/V_0\right)^n \ {\rm with} \ n = E/(h \nu),
\]
(in modern notation), where $E$ is the total energy.  This is of the
same form as that of a gas of $n$ particles.  From this remarkable
similarity in the two results, he concludes ``Monochromatic radiation
of small energy density behaves, as long as Wien's radiation law is
valid, for thermodynamic considerations, as if it consisted of
mutually independent energy quanta of magnitude $R\beta\nu/N$''.  (The
quantity $R\beta\nu/N$ is now denoted by $h \nu$).  This was the
introduction by Einstein of light quanta hypothesis. 

In the light quantum picture of Einstein ``in the propagation of a
light ray emitted from a point source, the energy is not distributed
continuously over ever-increasing volumes of space, but consists of a
finite number of energy quanta localised at points of space that move
without dividing, and can be absorbed or generated as complete
units''.  He then went on to apply the light quantum hypothesis to
other phenomena involving the generation and transformation of light.
The most important of these was his treatment of photoelectric
effect.  They also involved his successful application to eluciding
the Stokes' rule in photoluminescence and to the ioninsation of a gas
by ultraviolet light.
\bigskip

\noindent {\large 3.2. \underbar{The Photoelectric Effect:}}
\medskip

In 1887 Heinrich Hertz observed that the ultraviolet light incident on
metals can cause electric sparks.  In 1899 J.J. Thomson established
that the sparks are due to emission of the electrons.  Phillip Lenard
showed in 1902 that this phenomenon, now called the Photoelectric
effect, showed ``not the slightest dependence on the light intensity''
even when it was varied even a thousandfold.  He also made a
qualitative observation that as phtoelectron energies increased with
the increasing light frequency.  The observations of Lenard were hard
to explain on the basis of electromagnetic wave theory of light.  The
wave theory would predict an increase in photoelectron energy with
increasing incident light intensity and no effect due to increase of
frequency of incident light.

On the Einstein's light quantum picture, a light quantum, with energy
$h\nu$, on colliding with an electron in the metal, gives it's entire energy
to it.  An electron from the interior of a metal has to do some work,
$W$, to escape from the interior to the surface.  We therefore get the
Einstein phtoelectric equation, for the energy of the electron $E$,
\[
E = h\nu - W.
\]
Of course electron may loose some energy to other atoms before
escaping to the surface, so this expression gives only the maximum of
phto-electron energy which would be observed.  One can see that
Einstein's light quantum picture explains quite naturally the
intensity independence of photoelectron energies and gives a precise
quantitative prediction for it's dependence on incident light
frequency.  It also predicts that no photoelectrons would be observed
if $\nu < \nu_0$ where $h\nu_0 = W$.  The effect of increasing light
intensity should be an increase in the number of emitted electrons and
not on their energy.  Abram Pais has called this equation as the second
coming of the Planck's constant.

Robert A. Millikan spent some ten years testing Einstein equation and
he did the most exacting experiments.  He summarised his conclusions
as well as his personal dislike of light quantum concept, as follows:
``Einstein's photoelectric equation $\cdots$ appears in every case to
predict exactly the observed results $\cdots$ yet the semi-corpuscular
theory by which Einstein arrived at his equations seems at present
wholly untenable'' (1915) and ``the bold, not to say reckless
hypothesis of electromagnetic light corpuscle'' (1916).
\bigskip

\noindent {\large 3.3. \underbar{Envoi}}
\medskip

Einstein's light quantum paper, which was titled, \"Uber einen die
Erzeugung und Verwandlung des Lichtes betreffenden heuristichen
Geischtpunkt'' (on a heuristic point of view concerning the generation
and transformation of light), was completed on March 7, 1905 and
appeared in Annalen der Physik \underbar{17}, 132-148 (1905) and was
received by them on March 18, 1905.

It was thus his first paper during his annus mirabilis during which
he also wrote papers on Brownian motion, special theory of
relativity, and $E = mc^2$.  Though in public mind he is associated
indissolubly with relativity, with relativity as his most
revolutionary contribution, Einstein himself regarded his light
quantum paper among his papers written in 1905 as the ``most
revolutionary''. The opinion of the recent historians of science is
tending to agree with Einstein about it.  He was awarded Nobel prize
for 1921 in Physics for this paper which was announced in Nov. 1922.
Paranthetically his Nobel Lecture is on relativity theory.

Einstein's light-quantum is now known as ``photon'', a name given by
G.N. Lewis in as late as 1926.  Though Einstein talked about photon
energy $E = h\nu$, it is curious that he introduced the concept of
photon momentum $\vec p$, with magnitude $|\vec p| = h\nu/c$ only in
1917.  As we have seen even Millikan did not believe in photon concept
in 1915-16 despite his having spent years on experimental work
confirming it.  In 1923, the kinematics of the Compton effect was
worked out on the basis of it being an elastic electron-photon
scattering by A.H. Compton.  After that it was generally accepted by
physicists that light sometimes behaves as a photon.
\bigskip

\noindent {\Large\bf 4. Contributions to the Old Quantum Theory}
\medskip

\noindent {\large 4.1. \underbar{Specific Heat of Solids:}}
\medskip

Both Planck in 1900, and Einstein 1905 used the quantum theory to
understand problems of radiation.  Einstein in 1907 was first to apply
it to the problems of matter.  This was the problem of specific heat
of solids.

In 1819 Pierre Dulong and Alexis Petit, as a result of their joint
experimental work on a number of metals and sulpher at room
temperature, noted that all of them have almost the same specific heat
$C_V$, at constant volume, with a value of 6 calories per mol. per
$^{^\circ}K$ i.e. $C_V = 3R$.  Here $R$ is universal gas constant.
When other solids were investigated, especially carbon, the deviations
were found from the Dulong-Petit Rule.  In early 1870's Friedrich
Weber conjectured and then verified that $C_V$ approaches the value
$3R$ even for those cases at higher temperature i.e. $C_V = 3R$ is
only an asymptotic result.  Theoretically Ludwig Boltzmann applied
energy equipartition theorem to a three dimensional lattice crystal
and showed that $C_V = 3R$.  However the generality of the theorem left no
scope for any deviations from this result within classical physics.
There were similar problems which arose in the application of energy
equipartition theorem for gases.  As Lord Rayleigh noted in 1900
``What would appear to be wanted is some escape from the destructive
simplicity of the general conclusions (following from energy
equipartition theorem)''.  As we have noted earlier Lord Kelvin
regarded this problem as one of the clouds on the horizon of classical
physics.  

Einstein was first to realise that a use of equipartition theorem of
classical statistics leads to Rayleigh-Jeans radiation law which was
only asymptotically correct for large temperature.  To get the correct
Planck's radiation law one had to use quantum theory.  It was
therefore natural for him to try the same remedy to the problem of
specific heat of solids.  Besides he was always inclined to a
symmetrical treatment of radiation and matter.

Einstein assumed a simple model of the solid.  It is that of three
dimensional crystal lattice where all the atoms on the lattice oscillate
harmonically and independently and with the same frequency.  For a
solid with $N$ atoms we thus have a system of $3N$ harmonic
oscillators of frequency $\nu$.  We thus have using the earlier
expression, used in deriving Planck's expression for the average
energy of an oscillator of frequency $\nu$, and in thermal equilibrium
at temperature $T$, we get for the total energy $U$ of the solid,
\[
U = 3N \cdot {h\nu \over e^{h\nu/kT} - 1}.
\]
This leads to Einstein's expression for specific heat for his model
\[
C_V = 3R {\xi^2 e^\xi \over (e^\xi - 1)^2}, \ \xi = {h\nu \over kT}.
\]
It has the desirable feature that for $\xi$ small i.e. large $T$, we
get the Dulong-Petit result i.e.
\[
C_V \longrightarrow 3R \ {\rm as} \ \xi \rightarrow 0,
\]
which is the classical equipartion result.  It provides a one
parameter, i.e. $\nu$, formula for the specific heat of a solid.  The
deviations from Dulong-Petit value are also in broad agreement with
the experimental data.  The model of solid assumed is too simplistic
in that only a single frequency is assumed for all the oscillations.
It was improved by Peter Debye in 1912, and a more exact treatment of
atomic oscillations was given by Max Born and Theodore von K\'arm\'an
in 1912-1913.

A preliminary formulation of the third law of thermodynamics was given
by Walter Nernst in Dec. 1905 according to which the entropy of a
system goes to zero at $T = 0$.  Einstein's specific heat expression
has the property that $C_V \rightarrow 0$ as $T \rightarrow 0$ and
provides the first example of a model which is consistent with
Nernst's heat theorem, as was noted by Nernst in 1910.
\bigskip 

\noindent {\large 4.2. \underbar{Wave-Particle Duality}}
\medskip

In his 1905 Einstein had used phenomenological Wien's radiation law to
argue the particle nature of light.  In 1909 he used Planck's
radiation law to argue that light has both a particle and a wave
aspect.  For this purpose he calculated an expression for mean of
square of energy fluctuations $\langle \epsilon^2 (\nu,T)\rangle$ in
the frequency interval $\nu$ and $\nu + d\nu$.  From general
thermodynamic considerations, we have
\[
\langle \epsilon^2 (\nu,T)\rangle = kT^2 v d \nu {\partial \rho
(\nu,T) \over \partial T},
\]
in a subvolume $v$.

If we calculate this quantity using Rayleigh-Jeans radiation law $\rho
= \rho_{R-J} (\nu,T)$, we obtain
\[
\langle \epsilon^2(\nu,T)\rangle_{R-J} = {e^2 \over 8\pi\nu^2}
\rho^2_{R-J} v d \nu.
\]
Note that Rayleigh-Jean derivation is based on wave picture of light.
If on the other had we calculate this quantity using Wien's radiation
law, $\rho = \rho_{\rm Wien} (\nu,T)$, we obtain
\[
\langle \epsilon^2(\nu,T)\rangle_{\rm Wien} = h \nu \rho_{\rm Wien}
vd\nu. 
\]
As we know Wien's radiation law support a particle picture of light.

We now use the correct Planck's law of radiation $\rho = \rho_{\rm
Planck} (\nu,T)$ and obtain
\[
\langle \epsilon^2(\nu,T)\rangle_{\rm Planck} = h\nu\rho_{\rm Planck}
vd\nu + {c^2 \over 8\pi\nu^2} \rho^2_{\rm Planck} vd\nu.
\]
It is a very suggestive expression.  The first term is of the form we
get using Wien's law and supporting the particle picture light, while
the second term has the same form as that given by Rayleigh-Jeans law
which uses a wave picture of light.  We also know that the
contribution to the mean square fluctuations arising from independent
causes are additive.  This radiation has both wave and particle
aspects.  This was the first appearance in physics of wave-particle
duality, here for light radiation.

Einstein was quite prophetic in his remarks on the implications of
these results.  He says ``it is my opinion that the next phase in the
development of theoretical physics will bring us a theory of light
which can be interpreted as a kind of fusion of the wave and emission
theory $\cdots$ wave structure and quantum structure $\cdots$ are not
to be considered as mutually incompatible $\cdots$.  We will have to
modify our current theories, not to abandom them completely''.
\bigskip

\noindent {\large 4.3. \underbar{Einstein's A and B coefficients and
the discovery of stimulated Emission} \\ \underbar{of Light}}
\medskip

In 1916-1917 Einstein gave a new and wonderful derivation of Planck's
radiation law which provides a lot of new insights.  As he wrote to his
friend Michel Besso, in 1916, ``A splendid light has dawned on me
about the absorption and emission of radiation''.

He considers the thermodynamic equilibrium of a system comprising a
gas of ``molecules'' and radiation.  The ``molecules'' here refers to
any material system which is interacting with radiation.  Let the
energy levels of the ``molecules'' by denoted by $E_m$ and let the
number of ``molecules'' be given by $N_m$ when they occupy the energy
level $E_m$.

Consider two of these levels $E_2$ and $E_1$ with $E_2 > E_1$ and
consider the transitions from level 2 to level 1 and the reverse.
Einstein postulates that the number of transitions, in time $dt$, in
the ``molecules'' for the higher state $E_2$ to the lower state $E_1$
consists of two components.  One of these due to spontaneous jumps
from $E_2$ to $E_1$.  The number of transition however is given by the
term $A_{21} N_2 dt$.  Here the coefficient $A_{21}$ is related to the
intrinsic probability ofthis jump and does not depend on the radiation
density.  The second of these is due to stimulated emission of
radiation.  The number of transitions is here taken to be given by the
term $B_{21} N_2 \rho dt$ and is taken proportional to the radiation
density $\rho$.  Here the coefficient $B-{21}$ is related to the
probability of this process. The presence of radiation will also
induce transitions from the lower level 1 to higher level 2.  The
number of these transitions is taken to $B_{12} N_1 \rho dt$ and is
again taken proportional to the radiation density $\rho$.  The
coefficient $B_{12}$ again is related to the probability of this
process.  The $A_{ij}$'s and $B_{ij}$'s are called Einstein's $A$ and
$B$ coefficients.

In equilibrium the number of transitions from level 1 to level 2 must
be same as the number of transitions from level 2 to level 1.  We
therefore get the relation
\[
N_2 (A_{21} + B_{21} \rho) = N_1 B_{12} \rho.
\]
or 
\[
\rho = {(A_{21}/B_{21}) \over \left({B_{12} \over B_{21}}\right)
\left({N_2 \over N_1}\right) - 1}.
\]
Following Boltzmann, we have
\[
N_m = p_m e^{-E_m/kT},
\]
where $p_m$ is the relevant weight factor, and using it, we get
\[
\rho = {(A_{21}/B_{21}) \over \left({B_{12} p_2 \over B_{21}
p_1}\right) e^{(E_1-E_2)/kT} - 1}.
\]
From Wiens displacement we conclude that
\[
E_2 - E_1 = h\nu,
\]
a relation given by Bohr in 1913.  These transitions must involve
emission or absorption of radiation of frequency $\nu$.  Further for
large temperatures, i.e. $T \rightarrow \infty$, the $\rho$ must reduce to
Rayleigh-Jean's law.  This is possible only if we have
\[
{A_{21} \over B_{21}} = {8\pi h \nu^3 \over c^3}
\]
\[
p_2 B_{12} = p_1 B_{21}.
\]
Through this analysis we have got insights into the probabilities of
transitions and correct quantitative relations between them.  A
calculation of these was not possible until full apparatus of quantum
electrodynamics was in place which came much later, only in 1927.

The concept of stimulated emission, given by the coefficient $B_{21}$,
was introduced by Einstein here for the first time.  He was forced to
this step, since otherwise he would wave have been led to Wien's
radiation law by these considerations and not to the correct Planck's
law.  This concept is of fundamental importance in the theory of
lasers. 
\bigskip

\noindent {\Large\bf 5. Quantum Statistics : Bose and Einstein}
\medskip

The last great contribution to quantum theory, before the advent of
quantum mechanics, by Einstein was to develop quantum statistics for a
system of material particles.  Here the original idea was due to the
Indian physicist.  Satyendranath Bose from Dacca University and was
given in the context of radiation theory.  Einstein extended it to
matter.  As such this quantum statistical method is known as Bose
Statistics or Bose-Einstein statistics.  All integral spin particles
in the nature have been found to obey this statistics and are called
``Bosons''.  All half-odd integral spin particles obey Fermi-Dirac
statistics, which was given later in 1926 and are called ``Fermions''.  
\bigskip

\noindent {\large 5.1. \underbar{Bose:}}
\medskip

On June 4, 1924 Bose sent a short paper to Einstein containing a new
derivation of Planck's law.  It was accompanied by a very unusual
request to translate it into German and get it published in
Zeitschrift f\"ur Physik, if he found it worthwhile.  Bose explained
his chutzpah in doing it by saying ``Though a complete stranger to
you, I do not feel any hesitation in making such a request, because we
are all your pupils though profiting only by your teachings through
your writings''.  He also mentioned that he ``was the one who
translated your paper on Generalised Relativity'' when the first ever
english translation of the relativity papers of Einstein was published
by the Calcutta University in 1920.  We also know now, through William
Blanpied, that this paper had earlier been rejected for publication by
the Philosophical Magazine.

Bose noted ``since it's (Planck's law's) publication in 1901, many
methods for deriving this law have been proposed $\cdots$.  In all
cases it appears to me that the derivations have not been sufficiently 
justified from a logical point of view.  As opposed to these, the
light quantum combined with statistical mechanics (as formulated to
meet the needs of the quantum) appears sufficient for the derivation
of the law independent of the classical theory''.

Bose's idea was to regard the blackbody radiation as a free photon gas
and then treat it by the method of statistical mechanics.  This was
his strategy to derive Planck's radiation law in a logically
consistent manner.

Now photons of frequency $\nu$ have energy $h\nu$ and a momentum, with
magnitude $ph\nu/c$, on the light quantum hypothesis of Einstein.  A
straightforward calculation of the phase space volume element leads to
the factor $4\pi p^2 dp V$, where $V$ is the volume of the gas.  Bose
multiplied it by a further factor of 2, in order to take into account
the two polarisation states of the light, to obtain $8\pi p^2 dp V$.
If we now divide it by a factor $h^3$, following Planck's proposal of 1913
``that phase space cells have a volume $h^3$'' we obtain for the
number of phase space cells in this phase space volume element $8\pi
p^2 dp V/h^3$.  This leads to, using $p = h\nu/c$, the first factor
$8\pi \nu^2 d\nu/c^3$ in the Planck's radiation law.  Bose has thus
shown that the number $A^S$ of the phase space cells between radiation
frequency $\nu^s$ and $\nu^s + d\nu^s$ to be given by 
\[
A^S = {8\pi(\nu^s)^2 V d\nu^s \over c^3}
\]
in a novel way.  Note that Bose obtained this factor here, unlike
Planck, without making any use of the electromagnetic theory.  Bose
emphasized this aspects of his derivation in his letter to Einstein.

If Bose had proceeded further and used the statistical methods of
Boltzmann, at this stage, he would have obtained Wien's law and not
the desired Planck's law.  He however chose to interpret $A^S$, not as
the number of ``particles'' but as number of ``cells'', which played
the role of ``particles'' in Boltzmann's counting.  This procedure
then leads to Planck's law.  This is equivalent to treating photons as
indistinguishable in contrast to classical Boltzmann statistics where
particles are identical but distinguishable.  To give a simple example
if we have to distribute two identical balls, which are
distinguishable, by being coloured red and blue, into three containers.
There are nine possible different configurations and probability of
each one is 1/9 (Boltzmann counting).  On the other had if two
identical balls are not distinguishable, as we are colour blind, then
there are only six possible different configurations.  This is so
since the red ball in one container and blue ball in other container
is indistinguishable from the configuration in which we interchange
the two balls.  The probability of each distinct configuration flow is
now 1/6 (Bose counting).
\bigskip

\noindent {\large 5.2. \underbar{Einstein:}}
\medskip

Einstein immediately saw the importance of the Bose's work and got it
published in Zeitschrift f\"ur Physik after translating it into German
together with an appreciative note.  Not only that, in view of his
predilection to treat radiation and matter on the same footing, he
extended it immediately to a gas of material particles during
1924-1925.  For a photon gas there is no constraint of holding the
total number of photons fixed but for material particles, let us say
``atoms'', we have also a new constraint to hold the total number
fixed.  This introduced another parameter, chemical potential, which
has to be determined using this constraint.  Bose had not commented on
the indistinguishably aspect in his paper.  To bring this aspect out
Einstein also rewrote the Bose's formula for the total number of
configuration in the form it is normally found in textbooks.

We have seen that Einstein's model of solids was the first known
example in which Nernst's theorem was valid.  The case of
Bose-Einstein gas, which Einstein worked out, provides first model of
a gas for which Nernst's theorem holds.

Einstein also studied the fluctuations for the ideal Bose-Einstein
gas, as he had done earlier for radiation.  On calculating the mean
square fluctuation $(\Delta n^2$ for the number $n(\epsilon)$ of atoms
having energy between $\epsilon$ and $\epsilon + d\epsilon$, he found
it to consist again of two terms 
\[
(\Delta n)^2 = n(\epsilon) + {n^2(\epsilon) \over Z(\epsilon)}
\]
where $Z(\epsilon)$ is the number of particles states in the energy
interval $\epsilon$ and $\epsilon + d\epsilon$.  The first term is the
expected one for particles.

For an interpretation of the second term, which implies a wave aspect
for matter, Einstein suggested that this is due to wave nature of
atoms as postulated by Louis de Broglie in his recent doctoral thesis of
1924.  Einstein was aware of this thesis as Pierre Langevin had sent
him a copy for his opinion, and it was only Einstein's favourable
comments on it which made Langevin accept de Broglie's thesis.
Einstein also suggested associating a scalar field with these waves.
\bigskip

\noindent {\large 5.3. \underbar{Bose-Einstein condensation}}
\medskip

A free boson gas undergoes a phase transition below a critical
temperature $T_{BE}$.  A macroscopic fraction of the atoms condense
into lowest energy state.  This phase transition is not due to
interparticle attractive interaction but is simply a manifestation of
the tendency of bosons to stick together.  This was again a first
solvable model for a phase-transition.

Despite lot of efforts it was not possible to experimentally test this
prediction of Bose-Einstein until quite late.  It was finally observed
only in 1995.  The Nobel Prize in Physics for the year 2001 was
awarded to Eric Cornell, Carl Wieman and Wolfgang Ketterle for this
discovery. 
\bigskip

\noindent {\Large\bf 6. Foundations of Quantum Mechanics}
\medskip

\noindent {\large 6.1. \underbar{Discovery of Quantum Mechanics}}
\medskip

After a quarter century of long and fruitful interaction between the
old quantum theory and the experimental work on atomic systems and
radiation, this heroic period came to an end in 1925 with the
discovery of Quantum mechanics.  It was discovered in two different
mathematical formulations viz first as Matrix Mechanics and a little
later as Wave Mechanics.

Werner Heisenberg discovered Matrix mechanics during April-June 1925.
A complete formulation was achieved by Max Born, Werner Heisenberg and
Pascual Jordan in October 1925.  After the mathematical formalism was
in place, the problems of it's interpretation arose.  At Copenhagen,
Niels Bohr and Heisenberg and others devoted their full attention to
this talk.  The resulting interpretation, called `The Copenhagen
Interpretation of Quantum Mechanics', was to dominate the physics,
despite some other contenders, for a long time.  Heisenberg proposed
his famous `uncertainty principle' in Feb. 1927 in this connection.
In this work 
he was strongly influenced by a conversation he had with Einstein in
1926 at Berlin.  Heisenberg acknowledged to Einstein the role which
relativity with it's analysis of physical observation had played in
his own discovery of matrix mechanics.  His motivation in formulating
it had been to rid the theory of physical unobservables.  Einstein
differed and said ``it is nonsense even if I had said so $\cdots$ on
principle it is quite wrong to try founding a theory on observables
alone $\cdots$.  It is the theory which decides what is observable''.

The second formulation, wave mechanics, was published during the first
half of 1926, as a series of four papers ``Quantization as an
Eigenvalue problem'' in Annalen der Physik by Erwin Schr\"odinger.  He
was led to study the papers of de Broglie, wherein he suggested that
matter should also exhibit a wave nature, through a study of
Einstein's papers on Bose-Einstein gas.  He preferred a wave theory
treatment to the photon treatment of Bose and avoid new statistics.
As he said ``That means nothing else but taking seriously the
de-Broglie-Einstein wave theory of moving particles'' in a paper on
Bose-Einstein gas theory.  His next step was to make the idea of
matter-waves more precise by writing a wave equation for them.  This
is the famous Schr\"odinger wave equation for matter waves resulting in
the birth of wave mechanics.  As Schr\"odinger acknowledged ``I have
recently shown that the Einstein gas theory can be founded on the
consideration of standing waves which obey the dispersion law of de
Broglie $\cdots$.  The above considerations about the atom could have
been presented as a generalisation of these considerations''.  As Pais
says ``Thus Einstein was not only one of three fathers of the quantum
theory but also the sole godfather of wave mechanics''.  The three
fathers alluded to here are Planck, Einstein and Bohr.  

The mathematical equivalence of these two formulation was soon
established by Schr\"odinger and Carl Eckart in 1927.

After the discovery of Quantum mechanics the focus of Einstein shifted
from applications of quantum theory to various physical phenomena to
the problems of understanding what the new mechanics means.  With his
deep committment to the reality of an objective world Einstein was not
in tune with the Copenhagen interpretation. 
\bigskip

\noindent {\large 6.2. \underbar{Discussions at Solvay Conferences}}
\medskip

The fifth Solvay Conference was held at Brussels in October 1927.  It
was in this meeting that the claim of completeness of quantum mechanics
as a physical theory was put forward first.  In this connection
Einstein discussed the example of single hole diffraction of the
electron in order to illustrate two contrasting points of view:
\begin{enumerate}
\item[{(i)}] ``the de Broglie-Schr\"odinger waves do not correspond to
a single electron but to a cloud of electrons extended in space.
The theory does not give any information about the individual
processes'', and 
\item[{(ii)}] ``the theory has the presentations to be a complete
theory of individual processes''.
\end{enumerate}

The first viewpoint is what is now known as Statistical or Ensemble
interpretation of quantum mechanics if we clarify the phrase ``a cloud
of electrons'' to refer to an ensemble of single electron systems
rather that to a many electron system.  This is the view which
Einstein held in his later work.  He was thus the originator of ``The
Statistical or Ensemble interpretation of Quantum mechanics''.  This
view was also subscribed to by many others including Karl Popper and
Blokhintsev.  It is essentially the minimalist interpretation of
quantum mechanics. 

The second view point is the one upheld by the Copenhagen School and
very many others and may be termed as the maximalist interpretation.
Here a pure state provides the fullest description of an individual
system e.g. an electron.

The setup envisaged by Einstein was as follows: Consider a small hole
in an opaque screen and let an electron beam fall on it from the left
side.  Let it be surrounded by
another screen, on the right side, a hemispherical photographic plate.
From quantum mechanics the probability of an electron hitting at any
point of the photographic is uniform.  In the actual experiment the
electron will be found to have been recorded at a single definite
point on the plate.  As Einstein noted that one has to ``presuppose a
very peculiar mechanism of action at a distance which would prevent 
the wave function, continuously distributed over space from acting at
two places of the screen simultaneously $\cdots$ if one works
exclusively with Schr\"odinger waves, the second interpretation of
$\psi$ in my opinion implies a contradiction with the relativity
principle''.  Here Einstein is worried about, what we now call ``the
collapse of the wave function'' postulate and it's consistency with
special theory of relativity.  Einstein therefore opted for the
statistical interpretation of Quantum Mechanics.  A detailed
discussion of this interpretation would be out of place here. 

Apart from the formal discussion remark of Einstein noted above there
were also lots of informal discussions between him Niels Bohr.  In
these discussions Einstein generally tried to evade or violate 
Heisenberg's
uncertainty relations for individual processes by imagining various
possible experimental setups and Bohr constantly trying to find the
reason as why they would not work.  The uncertainties involved were
taken to be due to errors involved in the simultaneous measurement of
position-momentum or energy-time pairs. These discussion continued
also at Solvay Conference held at 1930.  These dialogues are quite
famous and Niels Bohr wrote an elegant account of them later.  It is
generally agreed that in these discussions Bohr was successful in
convincing Einstein that it was not possible to evade the uncertainty
principle.  However later developments, such as Bohm's realistic model
have shown that these discussion are somewhat irrelevant to the
problem of interpretation of quantum mechanics.
\bigskip

\noindent {\large 6.3. \underbar{Quantum Nonseparability and
Einstein-Podolsky-Rosen Correlations}}
\medskip

In quantum mechanics if two systems have once interacted together and
later separated, no matter how far, they can not any more be assigned
separate state vectors.  Since physical interaction between two very
distant systems is neglegible, this situation is very
counterintuitive.  Schr\"odinger even emphasized this aspect, ``I
would not call that one but rather the characteristic of quantum
mechanics''.  More technically, this is so for all two particle system
having a nonseparable wave function.  A wave function is regarded as
nonseparable, if no matter what choice of basis for single particle
wave function is used, it cannot be written as a product of single
particle wave functions.  Such wave functions are called entangled.
The entanglement is a generic feature of two particle wave functions. 

In 1935, A. Einstein, B. Podolsky and B. Rosen (EPR) published a paper
``Can Quantum Mechanical Description of Reality be Considered
Complete?'' in Physical Review.  It had a rather unusual title for a
paper for this journal.  In view of this they provided the following
two definitions at the beginning of the paper: (1) A
\underbar{necessary} condition for the \underbar{completeness} of a
theory is that every element of the physical reality must have a
counterpart in the physical theory.  (2) A \underbar{sufficient}
condition to identify an element of reality: ``If, without in any way
disturbing a system, we can predict with certainty (ie with
probability equal to unity) the value of a physical quantity, then
there exists an element of physical reality corresponding to this
physical quantity''.

We now illustrate the use of these definitions for a single-particle
system.  Let the position and momentum observable of the particle be
denoted by $Q$ and $P$ respectively.  Since in an eigenstate of $Q$,
we can predict with certainty the value of $Q$, which is given by it's
eigenvalue in that eigenstate, it follows that the position $Q$ of the
particle is an element of physical reality (e.p.r.).  Similarly the
momentum $P$ is also an e.p.r.  The position $Q$ and the momentum $P$
however are not simultaneous e.p.r.  So at the single particle level
there is no problem with quantum mechanics, as far as these
definitions of `completeness' and `elements of reality' are concerned.

The interesting new things are however encountered when a two particle
system is considered.  Let the momenta and position of the two
particles be denoted respectively by $P_1$ and $Q_1$ for the first
particle and by $P_2$ and $Q_2$ for the second particle.  Consider now
the two-particle system in the eigenstate of the relative-position
operator, $Q_2-Q_1$ with eigenvalue $q_0$.  The relative position
$Q_2-Q_1$ can be predicted to have a value $q_0$ with probability one
in this state and thus qualifies to be an e.p.r.  We can also consider
an eigenstate of the total momentum operator, $P_1 + P_2$, with an
eigenvalue $p_0$.  The total momentum can be predicted to have a value
$p_0$ with probability one and thus also qualifies to be an e.p.r.
Furthermore relative position operator, $Q_2-Q_1$, and total momentum
operator, $P_1+P_2$, commute with each other and thus can have a
common eigenstate, and thus qualify to be \underbar{simultaneous}
elements of physical reality.

We consider the two-particle system in which two particles are flying
apart from each other having momenta in opposite directions and are
thus having a large spatial separation.  The separation will be taken
so that no physical signal can reach between them.  Let a measurement
of position be made on the first particle in the region $R_1$ and let
the result be $q_1$.  It follows from standard quantum mechanics that
instantaneously the particle 2, which is a spatially for away region
$R_2$, would be in an eigenstate $q_0 + q_1$ of $Q_2$.  The $Q_2$ is
thus an e.p.r. the position of second particle gets fixed to the value
$q_0 + q_1$ despite the fact that no signal can reach from region
$R_1$ to $R_2$ where the second particle is, a  ``spooky action at a
distance'' indeed.  On the other hand a measurement of the momentum
$P_1$ of the first particle, in the region $R_1$ can be carried out
and let it result in a measured value $p_1$.  It then follows from the
standard quantum mechanics, that the particle 2, in the region $R_2$
would be in an eigenstate of its momentum $P_2$ with and eigenvalue
$p_0 - p_1$.  The $p_2$ is thus also an e.p.r.  This however leads to
a contradiction since $Q_2$ and $P_2$ can not be a simultaneous
e.p.r. as they do not commute.  We quote the resulting conclusion
following from this argument as given by Einstein in 1949, \\
\underbar{EPR Theorem}: The following two assertions are not
compatible with each other \\ ``(1) the description by means of the
$\psi$-function is complete \\ (2) the real states of spatially
separated objects are independent of each other''.

The predilection of Einstein was that the second postulate, now
referred to as ``Einstein locality'' postulate, was true and thus EPR
theorem establishes the incompleteness of quantum mechanics. 

As Einstein said ``But on one supposition
we should in my opinion, absolutely hold fast: the real factual
situation of the system $S_2$ is independent of what is done, with
system $S_1$, which is spatially separated from the former.

Einstein, Podolsky and Rosen were aware of a way out of the above
theorem but they rejected it as unreasonable.  As they said ``Indeed
one would not arrive at our conclusion if one insisted that two or
more quantities can be regarded as simultaneous elements of reality
only when they can be simulateneously measured or predicted.  On this
point of view, either one or the other, but not both simultaneously,
of the quantities $P$ and $Q$ can be predicted, they are not
simultaneously real.  This makes the reality of $P$ and $Q$ depend
upon the process of measurement carried out on the first system, which
does not disturb the second system in any way.  No reasonable
definition of reality could be expected to permit this''.
\bigskip

\noindent {\large 6.4. \underbar{Later Developments}}
\medskip

David Bohm reformulated the Einstein-Podolsky-Rosen discussion in a
much simpler form in terms of two spin one-half particles in a singlet
state in 1951.  This reformulation was very useful to John Bell, who
in 1964, gave his now famous Bell-inequalities on spin correlation
coefficients following from Einstein locality for EPR correlations.
These inequalities are experimentally testable.  In experiments of in
increasingly higher precision and sophistication they have shown
agreement with quantum mechanics and a violation of local realism
though some loopholes remain.  Bell's work on hidden variable theories
and Einstein-Podolsky-Rosen correlations had a profound influence on
the field of foundations of quantum mechanics, in that it moved it
from a world of sterile philosophical discussions to a world of
laboratory experiments. 

More recently E.P.R. correlations and quantum entanglement has been
found useful in developing new technologies of quantum information
such as quantum cryphography, quantum teleportation.  They have ceased
to be embarrasments but are seen as useful resources provided by
quantum mechanics.  There are even hopes of developing quantum
computing which would be much more powerful that usual universal
Turing machines.

Einstein's legacy in physics still looms large.  Talking about his
work Max Born once said ``In my opinion he would be one of the
greatest theoretical physicists of all times even if he had not
written a single line of relativity''.

\newpage

\noindent {\Large\bf 7. Bibilographical Notes}
\medskip

A brief and nontechnical summary of this paper appeared in \\ Singh V.,
The Quantum Leap, Frontline, \underbar{22}, \#10, p. 22-24 (May 7-20,
2005). \\ It also overlaps in places with author's earlier related
writings cited in various notes below

\begin{enumerate}
\item[{1.}] On the life and science of Einstein, the literature is
enormous.  The best biography for physicists is, \\
Pais, A., `\underbar{Subtle is the Lord $\cdots$': The Science and the
Life of Albert Einstein}.  Clarendon Press, Oxford: Oxford University
Press, New York, 1982. \\ It however needs supplementing by more
recent work on foundational aspects of Quantum mechanics:  Also
important is \\ \underbar{Albert Einstein: Philospher Scientist} edited
by P.A. Schilpp, Library of Living Philosophers Vol. 7, La Salle,
Ill.: Open Court, 1949; (Harper Torch Books reprint, 1959). \\ It
contains Einstein's ``Autobiographical Notes'' and ``Reply to
criticism'' as well as a ``Bibilography of the writings of Albert Einstein
to May 1951'' apart from 25 papers on the Science and philosophy of
Einstein by various authors.  At a more popular level, we have \\
\underbar{Einstein: A Centenary Volume}, edited by A.P. French,
Harvard University Press, Cambridge, Mass. 1979; \\ Bernstein, J.,
\underbar{Einstein}, The Viking Press, New York, 1973.
\item[{2.}] (a) For the writings of the Einstein we have the
multivolume ongoing series, \\ \underbar{The Collected Papers of
Albert Einstein}, Princeton University Press, Princeton, N.J., 1987 --
..., \\ and the companion volumes \\ \underbar{The Collected Papers of
Albert Einstein: English Translation}, Princeton University Press,
Princeton, N.J., 1987 -- ... . \\ His papers from the miraculous year
1905 are available in english translation also in \\
\underbar{Einstein's Miraculous Year: Five Papers that changed the
Face of Physics} edited by John Stachel, Princeton, 1998 (Indian
reprint: Scientia, An imprint of Srishti Publishers for Centre for
Philosophy and Foundations of Science, New Delhi, 2001). \\ (b) for
the references to some specific Einstein papers discussed in this
paper, see \\ (i) {\it light quantum paper}: Annalen der Physik,
\underbar{17}, 132-148 (1905) \\ (ii) {\it specific heat of solids}:
Annalen der Physik, \underbar{22}, 180-190, 800 (1907) \\ (iii) {\it
wave-particle duality}: Phys. Zeitschrift, \underbar{10}, 185, 817
(1909) \\ (iv) {\it A- and B-coefficients}: Verh. Deutsch
Phys. Ges. \underbar{18}, 318 (1916); Mitt. Phy. Ges. (Zurich),
\underbar{16}, 47 (1916); Phys. Zeitschrift 18, 121 (1917).  \\ The
concept of photon momentum is introduced in the 1917 paper mentioned
here. \\ (v) {\it Bose-Einstein Gas}:
Sitzungber. Preus. Akad. Wiss. Phys. Math. Kl. p. 261 (1924); p. 3
(1925) and p. 18 (1928). \\ (vi) {\it Einstein-Bohr dialogues}: Bohr,
N., Discussions with Einstein on Epistemological Problems in Atomic
Physics, in the \underbar{Schilpp Volume} cited earlier. \\ (vii) {\it
E.P.R. Theorem}: Einstein, A., Podolsky, B., Rosen, N.,
Phys. Rev. \underbar{57}, 777 (1935), and \\ \underbar{Schilpp Volume} cited
earlier. \\ (c) Other english translations of his light quantum paper and
1917 paper on A-B coefficients are also available in \\ \underbar{The
World of Atoms}, edited by H.A. Boorse and L. Motz, Basic Books, New
York, 1966; and \\ D. ter Haar, \underbar{The Old Quantum Theory},
Pergamon, Oxford, 1967. \\ An english translation of Einstein's first
two paper on Bose-Einstein statistics is available also in \\
N.D. Sengupta, Phys. News \underbar{14}, \#1, 10 (March 1983) and
\underbar{14}, \#2, 36 (June 1983). \\ Einstein quotes used in the
text are from various sources and are sometimes slightly modified or
abridged. 
\item[{3.}] For Physics before Einstein, see \\ Whittaker, E.T.,
\underbar{A Theory of Aether and Electricity, Vol. 1, Classical
Theories}: Harper Torchbacks, 1960; \\ Bork, A.M., Science \underbar{152},
\#3722, 597, 29 April 1966; \\ Singh, V., Science Today, p. 19-23,
Jan. 1980.
\item[{4.}] For a historical account of Quantum Mechanics, see \\
Whittaker, E.T., \underbar{A Theory of Aether and Electricity,
Vol. 2.} Modern Theories (1900-1926), Harper Torchbacks, 1960; \\
Jammer, M., \underbar{The Conceptual Dvelopment of Quantum Mechanics},
New York, 1966; \\ Hermann, A., \underbar{The Genesis of Quantum
Theory (1899-1913)}, translated by C.W. Nash, MIT Press, Cambridge,
Mass. 1971, \\ Hund, F., \underbar{The History of Quantum Theory},
translated by G. Reece, Harrap, London, 1974; \\ Kragh, H.,
\underbar{Quantum Generations: A History of Physics in the Twentieth
Century}, \\ Princeton Univ. Press, 2001 (Indian reprint, Universities
Press, Hyderabad, 2001). 
\item[{5.}] The work of S.N. Bose appeared in \\ Bose, S.N.,
Zeits. fur Physik \underbar{26}, 178, 1924 and \underbar{27}, 384,
1924.  \\ Two english translations exist of both the papers, one by \\
Banerjee, B., Phys. News \underbar{5}, 2, 40, 42, June 1974 \\ and
another by \\ Theimer, O. and Ram, B., Am. J. Phys. \underbar{44},
1058 (1976) and \underbar{45}, 242, 1977. \\ About his life and
science, see \\ Blanpied, W.A., Am. J. Phys. \underbar{40}, 1212,
1972; \\ Singh, V., Science Today, p. 29-34, Jan. 1974; \\ Mehra, J.,
\underbar{Biographical Memories of the Fellows of Royal Society,
London}, \underbar{21}, 117, 1975; \\ Chatterjee, S.D.,
\underbar{Biographical Memories of the Fellows of Indian National} \\
\underbar{Science Academy}, \underbar{7}, 59, 1983.
\item[{6.}] For later developments on the foundations of quantum
mechanics see, \\ Ballentine, L.E., Rev. Mod. Phys. 42, 358, 1970 and
Am. J. Phys. \underbar{40}, 1763 (1972). \\ Jammer, M., \underbar{The
Philosophy of Quantum Mechanics}, New York, 1974; \\ Bell, J.,
\underbar{Speakable and Unspeakable in Quantum Mechanics}, Cambridge,
1987; \\ Selleri, F., \underbar{Quantum Mechanics versus Local
Realism: The Einstein-Podolsky-Rosen} \\ \underbar{Paradox}, Plenum,
1988; \\ Home, 
D., \underbar{Conceptual Foundations of Quantum Physics}, Plenum, New
York, 1997. \\ Nielsen, M. and Chuang, I.L., \underbar{Quantum
Computation and Quantum Information}, Camabridge 2000; \\ Singh, V.,
Quantum Mechanics and Reality, arXive: quant-ph/0412148, 2004; \\
Singh, V., Hidden variables, Noncontextuality and Einstein Locality in
Quantum Mechanics, arXive: quant-ph/0507182, 2005.
\end{enumerate}

\end{document}